\documentclass[amssymb,amsmath,superscriptaddress,footinbib]{revtex4}
\usepackage{amsmath,natbib,graphicx,subfigure,subeqnarray,fancyhdr,amsmath,multirow,color, mathrsfs,relsize}
\begin{document}

\renewcommand{\labelitemi}{$-$}
\newcommand{\change}[1]{{\color{black} #1}}
\newcommand{\Fc}{\mathcal{F}}\newcommand{\Rc}{\mathcal{R}}\newcommand{\dd}{\mathrm{d}}
\newcommand{\ee}{\mathrm{e}}\newcommand{\ci}{\mathrm{i}}\newcommand{\ib}{\mathbf{i}}
\newcommand{\jb}{\mathbf{j}}\newcommand{\kb}{\mathbf{k}}\newcommand{\ab}{\mathbf{a}}
\newcommand{\Fb}{\mathbf{F}}\newcommand{\fb}{\mathbf{f}}\newcommand{\Gb}{\mathbf{G}}
\newcommand{\Mb}{\mathbf{M}Ä}\newcommand{\nb}{\mathbf{n}}\newcommand{\Sb}{\mathbf{S}}
\newcommand{\Sbs}{\mathbf{S^*}}\newcommand{\Rb}{\mathbf{R}}\newcommand{\Sigb}{\boldsymbol{\Sigma}}\newcommand{\sigb}{\boldsymbol{\sigma}}
\newcommand{\Sigbs}{\boldsymbol{\Sigma^*}}\newcommand{\alphab}{\boldsymbol\alpha}
\newcommand{\omegab}{\boldsymbol{\omega}}
\newcommand{\epsb}{\boldsymbol{\epsilon}}
\newcommand{\sigmab}{\boldsymbol{\sigma}}
\newcommand{\tb}{\mathbf{t}}

\newcommand{\ub}{\mathbf{u}}
\newcommand{\xib}{\boldsymbol{\xi}}\newcommand{\eb}{\mathbf{e}}\newcommand{\vv}[1]{\underline{#1}}\newcommand{\ev}{\vv{e}}
\newcommand{\rv}{\vv{r}}\newcommand{\TT}[1]{\underline{\underline{#1}}}\newcommand{\omb}{\mathbf{\omega}}
\def\v{\vspace{3cm}}
\newcommand{\Ub}{\mathbf{U}}\newcommand{\xb}{\mathbf{x}}\newcommand{\rb}{\mathbf{r}}
\newcommand{\ssb}{\mathbf{s}}\newcommand{\Xb}{\mathbf{X}}\newcommand{\Pe}{\mbox{Pe}}\newcommand{\Da}{\mbox{Da}\,}
\newcommand{\mean}[1]{\left\langle #1\right\rangle}
\newcommand{\ddp}{[p]^\pm}\newcommand{\taub}{\mbox{\boldmath$\tau$}}\newcommand{\Fr}{\mbox{\textit{Fr}}}
\let\grad\nabla\newcommand{\z}{\zeta}\newcommand{\kk}{\kappa}\newcommand{\tkk}{\tilde{\kappa}}
\newcommand{\e}{\varepsilon}\newcommand{\zb}{\bar{\zeta}}\let\grad\nabla\let\bcdot\cdot
\newcommand{\half}{{\textstyle\frac{1}{2}}}
\newcommand{\textfrac}[2]{{\textstyle\frac{#1}{#2}}}
\newcommand{\LF}[1]{{#1}^{\mathrm{LF}}}\newcommand{\Lap}[1]{{#1}^{\mathrm{L}}}
\newcommand{\ds}{*\!*}\newcommand{\cond}[2]{\frac{\mathrm{D} #1}{\mathrm{D} #2}}
\newcommand{\pard}[2]{\frac{\partial #1}{\partial #2}}\newcommand{\totd}[2]{\frac{\mathrm{d}#1}{\mathrm{d}#2}}
\newcommand{\pardd}[3]{\frac{\partial^2 #1}{\partial #2 \partial #3}}
\newcommand{\Rey}{\mbox{Re}}\newcommand{\Imag}{\mbox{Im}}
\newcommand{\Fpint}{=\!\!\!\!\!\!\!\int}
\newcommand{\txi}{\tilde\xi}\newcommand{\dxi}{\delta\xi}
\newcommand{\tpsi}{\tilde\psi}\newcommand{\dpsi}{\delta\psi}
\makeatletter
\def\sgn{\mathop{\operator@font sgn}}
\makeatother
\allowdisplaybreaks[1]

\title{Wall-driven flow rate and its sensitivity in microfluidic channels: a reciprocal theorem}
\title{A reciprocal method to compute boundary-driven flow rates}
\title{A reciprocal theorem for wall-driven pumps}
\title{A reciprocal theorem for boundary-driven channel flows}
\author{S\'ebastien Michelin}
\email{sebastien.michelin@ladhyx.polytechnique.fr}
\affiliation{LadHyX -- D\'epartement de M\'ecanique, Ecole Polytechnique -- CNRS, 91128 Palaiseau, France.}
\author{Eric Lauga}
\email{e.lauga@damtp.cam.ac.uk}
\affiliation{Department of Applied Mathematics and Theoretical Physics, University of Cambridge, Cambridge CB3 0WA, United Kingdom.}
\date{\today}

\begin{abstract}
In a variety of physical situations, a bulk viscous flow is  induced by a distribution of surface velocities, for example in diffusiophoresis (as a result of chemical gradients) and above carpets of cilia (as a result of biological activity). When such boundary-driven flows are used to pump fluids, the  primary quantity of interest is the induced flow rate.  In this letter we propose a method, based on the reciprocal theorem of Stokes flows, to compute the net flow rate for  arbitrary flow distribution and periodic pump geometry  using solely stress information from a dual Poiseuille-like problem. After deriving the general result we apply it to straight channels of triangular, elliptic and rectangular geometries and quantify  the relationship between  bulk motion and surface forcing. 

\end{abstract}
\maketitle

The precise manipulation of micro- or nano-scale flows is essential to many recent developments and applications in microfluidics, in particular for biological analysis and screening~\citep{whitesides2006}. It has also long been recognized as important in  the biological world, where  flows are  used to transport, deliver, mix, and flush away nutrients, for example in mucus transport~\citep{sleigh1988} and in cytoplasmic streaming inside plant cells~\citep{goldstein2008microfluidics}. 
Generating a flow within  confined environment requires to overcome the resistance of viscous stresses either through some external or localized forcing. Hence, two broad classes of methods have been proposed in order to induce flow in soft and hard (MEMS) microfluidics, that relies either on mechanical  or on phoretic/osmotic forcing~\cite{ho98,beebe02,stonereview,squires2005,kirby_book}. 

Mechanically-driven flows are  most classically achieved in the lab by imposing a pressure difference between the inlet and outlet of the channel. Another example of mechanical actuation is given by ciliary carpets~\citep{brennen1977,gueron1999}, responsible for many biological functions  such as mucociliary flow in the lungs or egg transport down the oviduct of mammals~\citep{sleigh1988,halbert1976}, and recently reproduced artificially using magnetically-driven cilia~\citep{coq2011} or bacterial carpets~\citep{darnton04,kim06}. When the cilia layer is thin compared to the channel-size, its mechanical forcing on the flow effectively takes the form of a slip velocity along the channel wall. The concept is similar for cytoplasmic  flows inside plant cells, with the wall-driven motion being now induced by the motion of specialized molecular motors carrying cargos along the surface of the cells~\citep{goldstein2008microfluidics}.

In contrast to mechanical forcing, phoretic mechanisms  exploits the interaction of charged or neutral solute molecules with the solid boundaries and  externally- (or locally-) imposed concentration or temperature gradients, or electric fields. Interactions between solute molecules and solid boundaries are generally localized over nanometer-size layers near the walls due to the short-range nature of the interaction forces. For channels with cross-sectional length scales of microns or more, the phoretic forcing on the flow thus effectively results in a slip velocity at the wall~\cite{anderson1989,kirby_book,saville,michelin15}.

A variety of biological and physical situations of interest consists thus of confined geometries with prescribed flow velocities  at the wall. An important question and challenge is then to relate the prescribed boundary conditions to the resulting flow rate for a given geometry. This generally requires solving the incompressible Stokes equation within the channel for a prescribed boundary condition, which can lead to cumbersome algebra when the geometry or the distribution of flow forcing are not trivial~\cite{happel,KimKarrila,leal2007}.


Here, inspired by classical work   deriving  the locomotion  kinematics of  microswimmers driven by surface motion~\citep{stone96},  we show that the problem of computing the flow rate can be solved easily for periodic geometries for which the solution to a Poiseuille-like problem has already been found. We first present the derivation for an arbitrary geometry and  obtain explicitly the flow rate and its sensitivity to wall-forcing. Applying this result to the straightforward case of a circular channel leads to classical solutions, and we then extend the calculation to three straight channels of different cross-sectional geometries (elliptic, triangular and rectangular). These examples illustrate how our formal result allows us to compute the flow rate and its sensitivity directly, without explicitly solving for the entire flow field, and hence represents a useful tool for design and optimization of flow manipulation in microfluidic devices.

\begin{figure}
\begin{center}
\includegraphics[width=.6\textwidth]{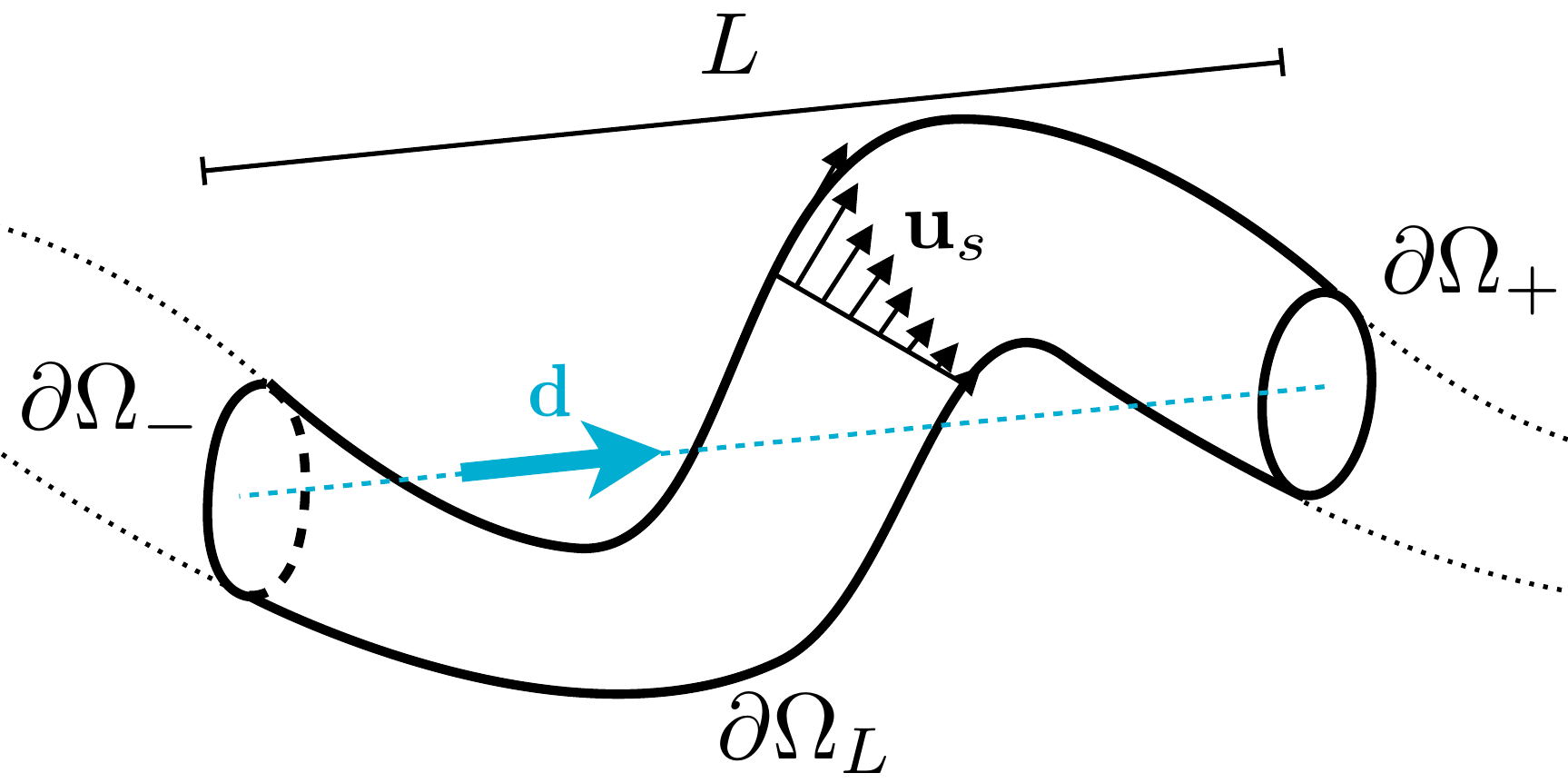}
\caption{Schematic representation of a periodic channel of period $L$. Along  one of the periods containing a  fluid volume $\Omega$, the lateral boundary of the channel is denoted  $\partial\Omega_L$. The  inlet and outlet have parallel surfaces 
 $\partial\Omega^-$ and $\partial\Omega^+$ and the unit vector joining these two surfaces, and perpendicular to them, is denoted $\bf d$.
}\label{fig:schema}
\end{center}
\end{figure}

The general problem addressed here is illustrated in Figure~\ref{fig:schema}.
We want to compute  net flow rate through a periodic channel resulting from a flow forcing at the walls. The periodic channel is obtained by the repetition of a pattern of arbitrary shape, and we denote by $\mathbf{d}$ the unit vector joining a given point of the inlet section to its periodic counterpart on the outlet cross-section of the pattern and $L$ the distance between them. 
Without any loss of generality we can assume that the inlet and outlet cross-sections $\partial\Omega^\pm$ are normal to $\mathbf{d}$. 
 Let $\Omega$ be the fluid volume contained in one period of the channel with boundary $\partial \Omega$, and $\partial\Omega=\partial\Omega_L\cup \partial\Omega^+\cup \partial\Omega^-$, with $\partial\Omega_L$ the lateral boundary of the channel. We seek to compute the volume flux $Q$ through the micro-channel defined as
\begin{equation}
Q=\int_{\partial\Omega^+}\ub\cdot\nb\dd S=\int_{\partial\Omega^-}\ub\cdot\nb\dd S,
\end{equation}
generated by a  prescribed forcing on $\partial\Omega_L$ in the form of net slip (tangential) velocity at the wall denoted $\ub^S$. 

Due to the linearity of  Stokes equations, the relationship between $Q$ and the forcing $\ub^S$ is linear, and it can be formally written as
\begin{equation}
Q=\int_{\partial\Omega_L}\mathbf{H}(\xb)\cdot\ub^S(\xb)\dd S,
\end{equation}
where $\mathbf{H}(\xb)$ is a (yet unknown) vector field defined on the lateral boundary that characterizes the {sensitivity} of the  flow rate in the channel to the boundary forcing at $\xb$ on $\partial \Omega_L$. The  direction of $\mathbf{H}$ indicates the direction of the most efficient forcing by a slip velocity, and  its intensity provides a map of the wall regions to which the flow rate is most sensitive.

A possible route to determine $\mathbf{H}(\xb)$ is to solve the Stokes flow problem resulting from the arbitrary surface field $\ub^S$ explicitly.
This boundary forcing generates a flow and pressure $(\ub,p)$ within the channel which satisfies the following Stokes problem
\begin{equation}\label{eq:stokesdirect}
\grad\cdot\sigmab=0,\qquad \nabla\cdot\ub=0,
\end{equation}
where $\sigmab = -p\mathbf{1}+2\eta\eb$  is the Newtonian stress tensor, $\eb$ the symmetric rate-of-strain tensor and $\eta$ the dynamic viscosity.
The associated  boundary conditions are
\begin{equation}
\ub=\ub_S\textrm{   on  }\partial\Omega_L,\qquad (\ub,p) \textrm{   periodic}.
\end{equation}
If an explicit solution  can be obtained, then the total volume flux within the channel and the flow rate sensitivity to wall forcing are easily obtained by integration along the inlet (or outlet) surface. Solving this Stokes flow problem for a single boundary forcing is however challenging  for complex geometries, and determining the solution for an arbitrary forcing is even more complex.

We propose here an  alternative approach. The idea  is to solve a simpler dual problem, once and for all, that does not depend on the boundary forcing and use its solution to obtain a direct relationship between  the distribution of slip velocity and the resulting flow rate. This approach makes an extensive use of the reciprocal theorem for Stokes  flows \cite{leal2007}.

The dual problem considered  is the periodic Stokes flow  in the same geometry forced by a constant volumetric forcing, i.e.
\begin{equation}\label{eq:stokesdual}
\nabla\cdot\sigmab^*=-G^*\mathbf{d},\qquad \nabla\cdot\ub^*=0,
\end{equation}
together with no-slip boundary condition, $\ub^*=0$, on $\partial\Omega$, and periodicity, i.e.~$\ub(\partial\Omega^+)=\ub(\partial\Omega^-)$ and $p(\partial\Omega^+)=p(\partial\Omega^-)$. Here, $\sigmab^*=-p^*\mathbf{1}+2\eta\eb^*$ denoted the Newtonian stress tensor associated with the velocity and pressure fields $(\ub^*,p^*)$  and $\eb^*$ the symmetric rate-of-strain tensor.

Multiplying Eq.~\eqref{eq:stokesdual} by $\ub$ and integrating over $\Omega$, after integration by parts, leads to
\begin{align}
-{G^*}\int_\Omega\ub\cdot\mathbf{d}\,\dd \Omega&=\int_{\partial\Omega}\ub\cdot\sigmab^*\cdot\nb\dd S-\int_\Omega \grad\ub:\sigmab^*\dd\Omega\nonumber\\
&=\int_{\partial\Omega}\ub\cdot\sigmab^*\cdot\nb\dd S-2\eta\int_\Omega \eb:\eb^*\dd \Omega.\label{8}
\end{align}
with  the unit normal $\nb$  pointing outside  the fluid domain. Following the classical approach from the reciprocal theorem, we multiply  Eq.~\eqref{eq:stokesdirect} by $\ub^*$ and integrate over $\Omega$, leading to
\begin{align}\label{9}
0=\int_\Omega\ub^*(\nabla\cdot\sigmab)\dd \Omega&=\int_{\partial\Omega}\ub^*\cdot\sigmab\cdot\nb\dd S-2\eta\int_\Omega\eb:\eb^*\dd\Omega,
\end{align}
where  we have used that   $\ub$ and $\ub^*$ are both divergence-free.
Equating the last terms of the right-hand sides of Eqs.~\eqref{8} and \eqref{9} leads to the relationship 
\begin{align}
\int_{\partial\Omega}\ub^*\cdot\sigmab\cdot\nb\dd S-\int_{\partial\Omega}\ub\cdot\sigmab^*\cdot\nb\dd S&={G^*}\int_\Omega\ub\cdot\mathbf{d}\,\dd\Omega.
\end{align}
The integral on the right-hand side of the equation can be simplified by noting that 
\begin{equation}
\int_\Omega\ub\cdot\mathbf{d}\,\dd\Omega= QL,
\end{equation}
where $Q$ is the volume flux through the channel, independent of the cross-section used to compute it, and of its orientation. Due to the periodicity of the problem, the integrals on $\partial \Omega_+$ and $\partial\Omega_-$ cancel each other out, and using the boundary conditions for $\ub$ and $\ub^*$ on $\partial \Omega_L$ we obtain the final result for the flow rate as
\begin{equation}\label{eq:result}
Q=-\frac{1}{L G^*}\int_{\partial\Omega_L}\ub_S\cdot\sigmab^*\cdot\nb\,\dd S.
\end{equation}
\change{The dual problem is linear; therefore, $\sigmab^*$ is proportional to $G^*$ and $Q$ does not depend on this arbitrary constant, which is kept here nonetheless for dimensional consistency of the equations.}

The result in Eq.~\eqref{eq:result}  provides thus an explicit expression for the sensitivity of  the flow rate  to wall forcing, $\mathbf{H}(\xb)$, in terms of the stress tensor of the dual problem
\begin{equation}
\mathbf{H}(\xb)=-\frac{\sigmab^*\cdot\nb}{L G^*}\cdot
\end{equation}
From this result, one is able to compute the average flow rate within the channel for {arbitrary} slip velocity at the wall. Note that the previous equation was  obtained for a purely tangential flow forcing at the wall (i.e.~$\ub_S\cdot\nb=0$), in order to ensure that the flow rate $Q$ is independent of the cross section considered. In that case, only the deviatoric part of the viscous stress tensor is needed to compute $\mathbf{H}$. The approach may however be generalized for flow forcing including a normal component by replacing $Q$ with its spatial average. 
Furthermore, the derivation was done is three dimensions but is directly applicable to two-dimensional channels by applying our result to a slice of the two-dimensional channel since the invariance in the third dimension guarantees the cancellation of the integrals on the lateral surfaces.


In the case of a circular cylindrical channel of axis $\eb_z$ and radius $R$, the dual problem Eq.~\eqref{eq:stokesdual} is  the classical Poiseuille flow. In that case, the velocity field and stress tensor are obtained in cylindrical polar coordinates as
\begin{equation}
\ub^*=\frac{G^*}{4\eta}(R^2-r^2)\eb_z,\qquad 2\eta \mathbf{e}^*=-\frac{G^*r}{2}(\eb_z\eb_r+\eb_r\eb_z).
\end{equation}
Here, the surface traction $\sigmab^*\cdot\nb$ on the boundary is uniform, which is expected by symmetry. The flow rate created by a localized forcing on the boundary is thus independent of the position of that forcing location and the sensitivity $\mathbf{H}(\xb)$ is therefore uniform, with value
\begin{equation}
\mathbf{H}(\xb)=\frac{R}{2L}\eb_z.
\end{equation}
Hence, for any distribution of velocity $\mathbf{u}_b(\theta,z)$ along the channel, and as expected, the channel flow rate is simply the average of the boundary velocity
\begin{equation}
Q=\frac{R}{2L}\int_{\partial\Omega_L}\mathbf{u}_b\cdot\eb_z\dd S=\pi R^2\langle u_{bz}\rangle.
\end{equation}

Turning now to more complex geometries, the reciprocal theorem approach is particularly  powerful when a solution of the dual problem can be obtained easily. We highlight three of such solutions for straight channels of axis $\eb_z$.  

We first consider a straight channel of elliptical cross-section with major and minor axes denoted $a$ and $b<a$, respectively.  The dual flow problem can  be solved analytically and we have
\begin{align}
\ub^*=&\frac{a^2b^2G^*}{2\eta(a^2+b^2)}\left(1-\frac{x^2}{a^2}-\frac{y^2}{b^2}\right)\eb_z,\\
2\eta \eb^*=&-\frac{G^*}{(a^2+b^2)}\left[b^2x(\eb_z\eb_x+\eb_x\eb_z)+a^2y(\eb_z\eb_y+\eb_y\eb_z)\right].
\end{align}

The contour of the elliptical cross-section can be parameterized as
$x=a\cos \theta,\, y=b\sin \theta,$
with $0\leq \theta\leq 2\pi$. We are  interested in the flow resulting from tangential boundary forcing $\ub_b(\theta,z)=u_{b}(\theta,z)\eb_z$ along the axis of the pipe. The local tangent and normal unit vectors in the cross section are 
\begin{align}
\boldsymbol\tau=\frac{-a\sin \theta\,\eb_x+b\cos \theta\,\eb_y}{\sqrt{a^2\sin^2\theta+b^2\cos^2\theta}}, \quad
\nb=\frac{b\cos \theta\,\eb_x+a\sin \theta\,\eb_y}{\sqrt{a^2\sin^2\theta+b^2\cos^2\theta}}\cdot
\end{align}
The sensitivity of the flow rate to forcing on the boundary is therefore given by 
\begin{equation}
\mathbf{H}(\xb)=\frac{ab\sqrt{b^2\cos^2\theta+a^2\sin^2\theta}}{L(a^2+b^2)}\,\eb_z.
\end{equation}
When the channel is non-circular, the sensitivity is not uniform anymore along the cross-section and is greatest in the regions of lowest curvature ($\theta=\pm\pi/2$, i.e.~along the $y$ axis). Instead, for an ellipsoidal cross-section with high aspect ratio, the sensitivity of the strongly-confined tips ($\theta=0,\pi$ i.e.~along the $x$ axis) is very small. 
Finally the flow rate is given by the integral
\begin{equation}
Q=\frac{ab}{(a^2+b^2)}\int_0^{2\pi} \langle u_b\rangle_z(\theta)(b^2\cos^2 \theta+a^2\sin^2 \theta)\dd \theta,
\end{equation}
where $\langle u_b\rangle_z$ the $z$-average of the boundary forcing.

Another example considers a  cylindrical channel with equilateral triangular cross-section of side $a$. We use cartesian coordinates so that the three sides have equations $y=0$  and $y\pm \sqrt{3}x={a\sqrt{3}}/{2}$.
Here again the dual flow problem is known analytically and given by
\begin{align}
\ub^*=&\frac{G^*y}{2\eta\sqrt{3}a}\left[\left(\frac{a\sqrt{3}}{2}-y\right)^2-3x^2\right]\eb_z.
\end{align}
On the bottom boundary $y=0$ the normal is $\nb = - \eb_y$, we have 
 $\sigma^*_{yz}=G^*\sqrt{3}/2a(a^2/4-x^2)$ 
 so the  sensitivity of the flow rate to a forcing on the bottom wall is then given by
\begin{equation}
\mathbf{H}(\xb)=\frac{\sqrt{3}}{2aL}\left(\frac{a^2}{4}-x^2\right) \eb_z.
\end{equation}
Clearly the sensitivity is non uniform in this case. The  values of $\mathbf{H}$ on the other two walls can be simply obtained by symmetry. A tangential forcing $\ub_b(x)=u_{b}(x)\eb_z$ applied on the bottom boundary therefore leads to a net flow rate
\begin{equation}
Q=\frac{\sqrt{3}}{2a}\int_{-a/2}^{a/2}\langle u_{b}\rangle _z(x)\left(\frac{a^2}{4}-x^2\right)\dd x.
\end{equation}
The  flow rate is more sensitive to a forcing in the central region of the channel walls than in one of the corners, a result that is expected due to the higher confinement and viscous friction in the corners of the triangle cross section. Obviously, the flow rate resulting from a forcing on each of the three edges can be obtained by superimposing the contribution of each side. Because the geometry is invariant by a rotation of $2\pi/3$ around the section's center, the sensitivity of the flow rate to boundary forcing is identical and parabolic on each side.

Another example considers a straight channel of cross-sectional dimensions $L_x$ and $L_y$, the geometry most widely used in microfluidics \cite{kirby_book}. Assuming a uniform problem along the channel axis $\eb_z$, the dual flow problem solution is known analytically in the form of a doubly infinite series and given by
\begin{align}
\ub^*=\frac{16G^*}{\eta\pi^4}&\mathlarger{\sum}_{p,q=0}^\infty\frac{\sin\left(\frac{(2p+1)\pi x}{L_x}\right)\sin\left(\frac{(2q+1)\pi y}{L_y}\right)}{(2p+1)(2q+1)\left[\left(\frac{2p+1}{L_x}\right)^2+\left(\frac{2q+1}{L_y}\right)^2\right]}\,\eb_z.
\end{align}
The sensitivity of the flow rate to wall forcing is then obtained for horizontal walls at $y=0$ and $y=L_y$  as
\begin{align}
\mathbf{H}_\textrm{hor}(x)&=\frac{16}{\pi^3LL_y}\mathlarger{\sum}_{p,q=0}^\infty\frac{\sin\left(\frac{(2p+1)\pi x}{L_x}\right)}{(2p+1)\left[\left(\frac{2p+1}{L_x}\right)^2+\left(\frac{2q+1}{L_y}\right)^2\right]}\,\eb_z,
\end{align}
while the sensitivity on the vertical walls $\textbf{H}_\textrm{vert}(y)$ at $x=0$ and $x=L_x$ can be obtained by symmetry.

\begin{figure}
\begin{center}
\includegraphics[width=.6\textwidth]{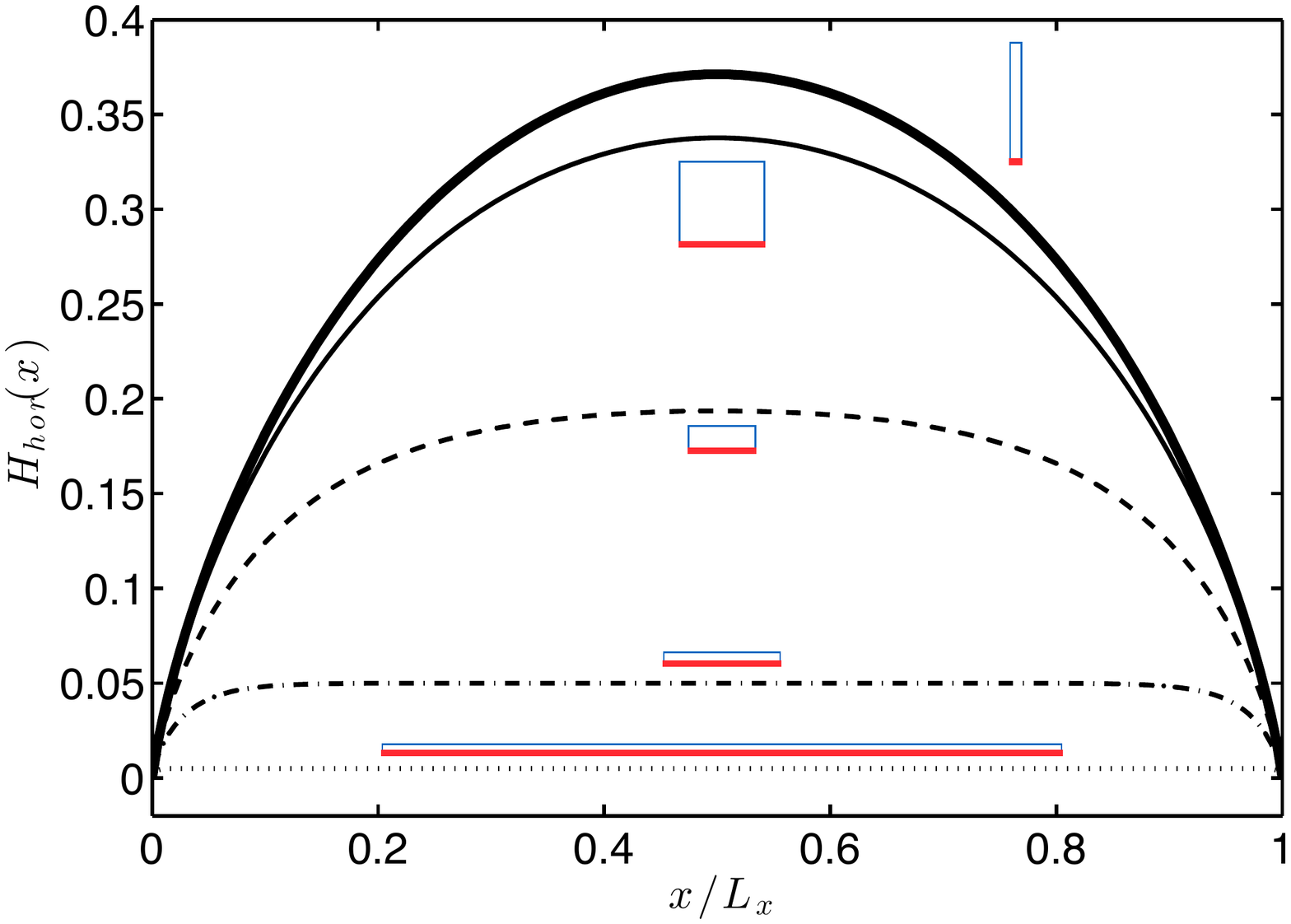}
\caption{Sensitivity of the flow rate to a slip forcing along the horizontal (bottom) wall of a rectangular channel for different aspect ratios: $L_y/L_x=0.01$ (dotted), $L_y/L_x=0.1$ (dash-dotted), $L_y/L_x=0.4$ (dashed), $L_y/L_x=1$ (thin solid) and $L_y/L_x=10$ (thick solid).}\label{fig:sensitivity_rectangle}
\end{center}
\end{figure}

The dependence of the sensitivity along the horizontal wall with the aspect ratio of the channel is shown on Figure~\ref{fig:sensitivity_rectangle}. In the limit of $L_y\ll L_x$, the microchannel is close to being two-dimensional and does not ``feel'' the presence of the vertical walls except within a neighborhood of size $O(L_y)$ near the edges, and thus 
the sensitivity  is approximately constant along the entire channel wall.
In the opposite limit ($L_y\gg L_x$), the flow resulting from a forcing on the bottom boundary ($y=0$) is not influenced by the presence of the upper wall and the sensitivity becomes independent of $L_y/L_x$. The sensitivity is maximum in the center of the channel as expected, as this is the location  minimizeing the effect of friction along the vertical walls.

The general framework presented in this Letter provides a direct method to compute the flow rate and sensitivity to wall-forcing within any periodic channel without actually solving the full Stokes flow problem for this particular forcing. Instead, provided that a dual Poiseuille-like solution is known for the same geometry, our main result, Eq.~\eqref{eq:result}, gives an explicit and direct expression of the net flow rate. Importantly, even in situations where the dual solution is not trivially obtained, it only needs to be derived once, be it analytically or numerically. Then, the flow rate's response to {any} wall forcing can be directly obtained from the integration of Eq.~\eqref{eq:result} which, in the cases where the dual solution is tedious, can be carried out computationally. Note that a similar result can be easily obtained when the stress (rather than the slip velocity) is prescribed at the boundary: in that case, the dual Poiseuille-like problem must be solved with zero-stress (rather than no-slip) boundary conditions. Then the sensitivity is proportional to the dual slip velocity and the flow rate is obtained as a boundary integral of the prescribed stress, similarly to Eq.~\eqref{eq:result}.

The original paper showing how to use the reciprocal theorem for low-\mbox{Re} propulsion~\citep{stone96} allowed to derive the velocity of a micro-swimmer in terms of the prescribed surface velocity boundary condition. Cases of interest there include  the envelope model of a ciliate (i.e.~when the cilia layer is  thin compared to the cell size so that the cilia beating can be replaced by a tangential slip velocity \citep{brennen1977}) or for phoretic swimmers which have generated much excitement recently as examples of fuel-based torque- and force-free artificial micro-swimmers~\citep{paxton2004,howse2007,golestanian2007}. \change{Such alternative approaches to solving the full Stokes flow problems around the moving swimmers have recently attracted much interest with generalizations to two-dimensional locomotion problems~\citep{elfring15}, multiple-body problems~\citep{papavassiliou2015} or Marangoni propulsion~\citep{masoud2014b}.}
The present paper can be very much seen as an extension of those ideas to wall-driven pumping, as it directly relates the flow rate to the kinematic forcing at the boundary of the channel. For example, our results  could be used to determine the net flow generated by biological or artificial  cilia carpets and analyze the influence of density, orientation and coordination of cilia. Also, the present paper could be an essential tool for designing diffusio-phoretic pumps, exploiting phoretic phenomena near active channel walls~\citep{michelin15}, and for analyzing how the flow rate of such pumps  depends on the detals of the  chemical boundary conditions.
\bigskip


Funding from the French Ministry of Defense  (DGA -- S.~M.) and the European Union (Marie Curie CIG -- E.~L.) is gratefully acknwoledged.


\end{document}